\documentclass{nature}

\usepackage[pdftex]{graphicx, xcolor}
\usepackage[pdftex]{hyperref}
\usepackage{siunitx}
\usepackage{amsmath,amssymb}
\usepackage{wasysym}
\usepackage{siunitx}
\usepackage[font=footnotesize,labelfont=bf]{caption}

\usepackage{graphicx}
\makeatletter
\let\saved@includegraphics\includegraphics
\AtBeginDocument{\let\includegraphics\saved@includegraphics}
\renewenvironment*{figure}{\@float{figure}}{\end@float}
\makeatother

\DeclareSIUnit\rpm{rpm}

\title{Granulation and suspension rheology: a unified treatment}

\author{Daniel J. M. Hodgson$^{1,*}$, Michiel Hermes$^2$, Elena Blanco$^{1}$ \& Wilson C. K. Poon$^1$}

\begin{document}

\maketitle

\begin{affiliations}
 \item SUPA and School of Physics and Astronomy, The University of Edinburgh, King's Buildings, Peter Guthrie Tait Road, Edinburgh, EH9 3FD, United Kingdom
 \item Soft Condensed Matter, Debye Institute for Nanomaterials Science, Utrecht University, Princetonplein 5, 3584 CC Utrecht, The Netherlands
\end{affiliations}


\begin{abstract}
Mixing a small amount of liquid into a powder can give rise to dry-looking granules; increasing the amount of liquid eventually produces a flowing suspension.  We perform experiments on these phenomena using Spheriglass, an industrially-realistic model powder.  Drawing on recent advances in understanding friction-induced shear thickening and jamming in suspensions, we offer a unified description of granulation and suspension rheology. A `liquid incorporation phase diagram' explains the existence of permanent and transient granules and the increase of granule size with liquid content. Our results point to rheology-based design principles for industrial granulation.
\end{abstract}


Incorporating a small amount of liquid into powders is a ubiquitous unit operation in industrial materials processing. In some cases, a minimal amount of liquid is used to produce matt solid granules. Such `wet granulation' \cite{Iveson2001} has long been used in the manufacturing of, e.g., detergents, drugs \cite{Salman2006} and gunpowder~\cite{marvin1875}. When too much liquid is added, the mixture becomes (in granulation jargon) `overwet' \cite{regime}: it turns into a flowing suspension and granulation fails. However, such high-solid-content dispersions are the desired end point for other applications, e.g., ceramic pastes, construction and medical cements, and molten chocolate \cite{blanco2019conching}. If insufficient liquid is added in these systems, flow fails -- they fracture and jam, e.g., near constrictions.

Applied research into these two areas has hitherto proceeded separately: where the interest of one community ends (`overwet'), the attention of the other begins. However, in terms of the amount of liquid incorporated, the preparation of granules and high-solid dispersions form a continuum, as is visually apparent in published snapshots taken from the process of `conching' chocolate \cite{blanco2019conching} (see also Supplementary Information \cite{SI}) and mixing concrete \cite{cazacliu2009concrete}. Thus, it is of interest to enquire whether, on some fundamental level, these two areas of industrial practice may be amenable to a single, unified description, with insights from each enriching the understanding of the other. Indeed, it has been suggested that the shear-induced jamming may be related to granulation \cite{Cates2005,Cates2014}: the fracturing of an `underwet dispersion' into granules is the system's response to the impossibility of homogeneous flow.

\begin{figure}
\begin{center}
\includegraphics[width=0.64\columnwidth]{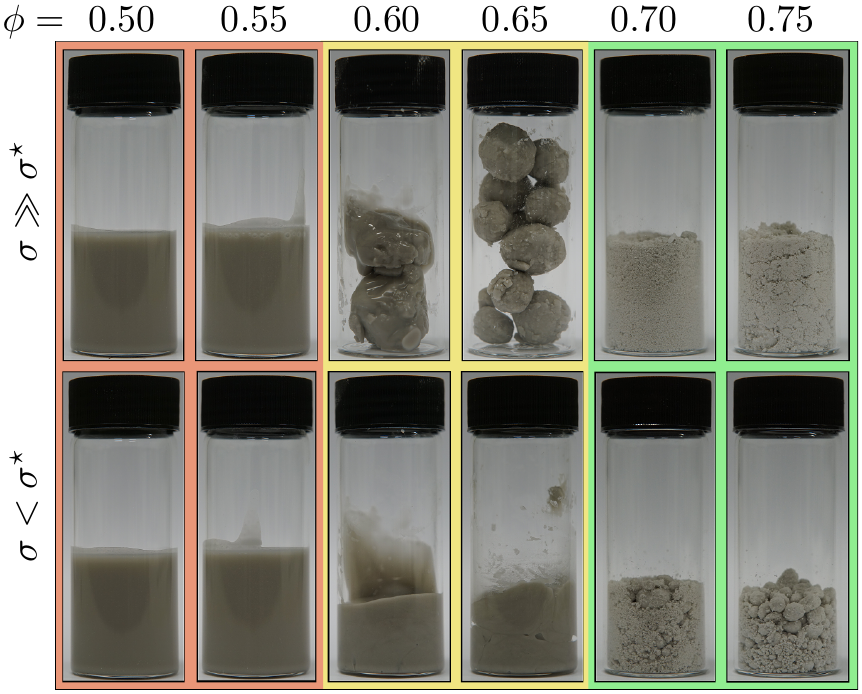}
\end{center}

\caption{The result of mixing at different volume fractions and stresses. (a) Output from the high-shear mixer, and (b) and subsequent low stress vortex mixing. Tube diameter = \SI{20}{\mm}. Colors correspond to the $\phi$ regimes in \autoref{fig:rheo}.
{\color[RGB]{233,150,122}$\blacksquare$} $\phi < \phi_\textrm{m}$: samples are flowing, liquid suspensions in both cases.  {\color[RGB]{230,240,130} $\blacksquare$} $\phi_\textrm{m} < \phi < \phi_\textrm{rcp}$: high-shear mixing produces solid-like granules, which melt to a flowing liquid suspension upon vortex mixing.
{\color[RGB]{144,238,144} $\blacksquare$} $\phi > \phi_\textrm{rcp}$: solid granules from the high-shear mixing remain solid upon vortex mixing.}
\label{fig:granulation}
\end{figure}

\begin{figure}
\begin{center}
\includegraphics[width=0.5\columnwidth]{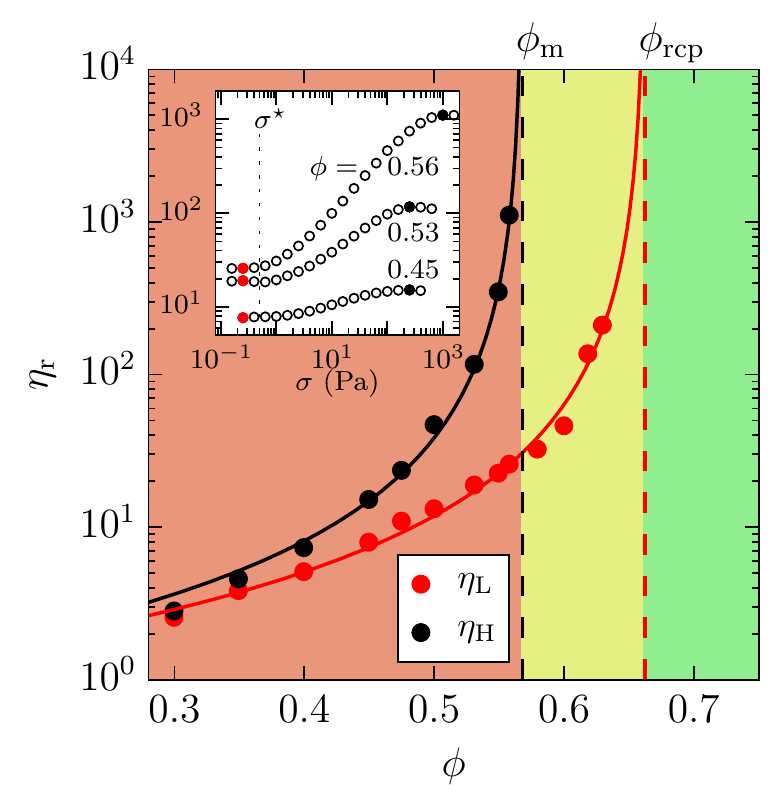}
\end{center}
\caption{Relative viscosity, $\eta_\textrm{r}$, as a function of volume fraction, $\phi$. The low-stress, $\eta_{\textrm{L}}$, ({\color{red} $\CIRCLE$}) and high-stress (or thickened) $\eta_{\textrm{H}}$, ($\CIRCLE$) branches with fits (solid lines) to \autoref{eq:KD} diverging at $\phi_\textrm{rcp}$ and $\phi_\textrm{m}$ respectively. {\color[RGB]{233,150,122} $\blacksquare$} $\phi < \phi_\textrm{m}$: the system flows at all applied stresses. {\color[RGB]{230,240,130} $\blacksquare$} $\phi_\textrm{m} < \phi < \phi_\textrm{rcp}$: the system can only flow for applied stresses $\lesssim \sigma^{\star}$.  {\color[RGB]{144,238,144} $\blacksquare$} $\phi \geq \phi_\textrm{rcp}$: samples never flow. \textit{Inset}: $\eta_\textrm{r}$ as a function of $\sigma$ for various $\phi$.  {\color{red} $\CIRCLE$}, $\CIRCLE$ = points plotted using the same symbols in the main graph.
\label{fig:rheo}
}
\end{figure}

\begin{figure}
\begin{center}
\includegraphics[width=0.5\columnwidth]{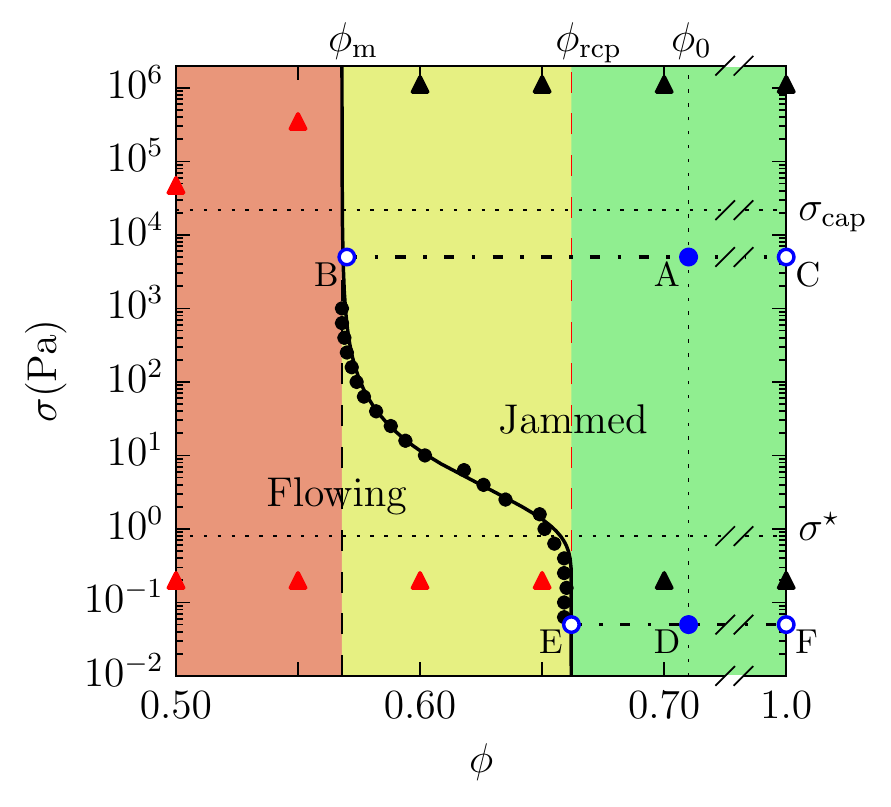}
\end{center}
\caption{A $\phi$-$\sigma$ phase diagram for liquid incorporation into powders. Bold curve and {$\bullet$}: $\sigma_\textrm{J}(\phi)$ derived from fitting the values of $\phi_{\textrm{J}}(\sigma)$ \cite{SI}. To its left and right, we find flowing and jammed states respectively. Color codling: as in Figs.~\ref{fig:granulation} and \ref{fig:rheo}. Mixing a sample with average composition $\phi_0$  at high stress (point $A$) gives jammed granules at composition $\phi_\textrm{m}$ and dry powder at composition $\phi = 1$, with relative proportions AC/AB. Mixing the same average composition at a lower stress (say, $\SI{0.05}{\pascal}$), $D$, gives granules at composition $E$ ($\phi = \phi_\textrm{rcp})$ and dry powder ($F$), with relative volumes of DE/DF.}
\label{fig:jamming}
\end{figure}

\begin{figure}[t]
\begin{center}
\includegraphics[width=0.5\columnwidth]{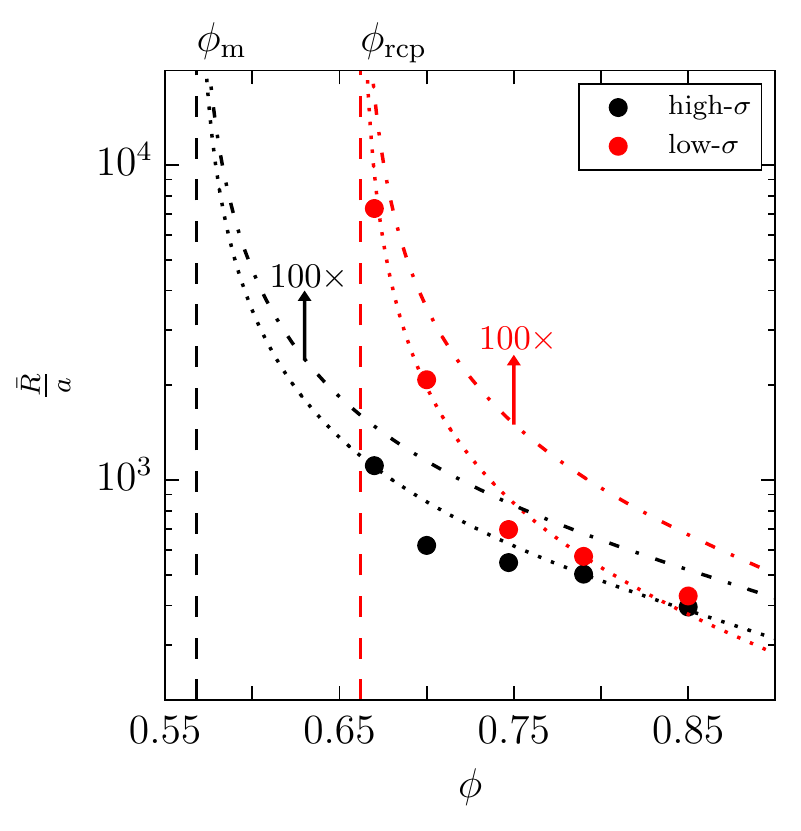}
\end{center}
\caption{The volume-weighted mean granule size, $\bar{R}$, in units of the mean primary particle radius, $a$, as a function of the average system composition, $\phi$; red: low stress, black: high stress. Lines: calculated $R/a$ for shell thickness $t_{\textrm{s}} = d$ in \autoref{eq:size} (dot-dashed, multiplied by 100 for comparison to data), and fitting \autoref{eq:size} to return $t_{\textrm{s}} = 74d$ and $56d$ for high and low stress data sets respectively (dotted).
\label{fig:size}
}
\end{figure}

We report an experimental study of the entire range of phenomena as progressively more liquid is incorporated into a model industrial powder of $\sim \SI{10}{\micro\meter}$ glass particles. We observe successively the formation of two different kinds of granules, which, at the point of `overwetting', merge into a high-solid-content dispersion, \autoref{fig:granulation}. We characterise the rheology of these dispersions, which shear thicken,  \autoref{fig:rheo}. Interpreting our data in the light of recent advances in suspension rheology \cite{Seto2013,Mari2014,wyart2014,Guy2015,Lin2015,comtet2017pairwise}, we construct a `liquid incorporation phase diagram', \autoref{fig:jamming}, which unifies the various possible states into a single conceptual scheme arising from shear-induced jamming. A `tie line construction' satisfying the `lever rule' explains the dependence of granule size on solid content, \autoref{fig:size}. Our results confirm theoretical proposals \cite{Cates2005,Cates2014} that link suspension rheology and wet granulation.

Suspensions of glass spheres in a mixture of glycerol and water at various solid volume fractions, $\phi$, were mixed using different methods to apply `high' and `low' stresses (see \textit{Methods} for details). Note that $\phi = V_{\textrm{solid}}/(V_{\textrm{solid}} + V_{\textrm{liquid}})$, so that in a dry powder $\phi=1$.


The result of high and low stress mixing at $0.5 \leq \phi \leq 0.75$ is shown in \autoref{fig:granulation}, where we have color coded different regimes. Samples with $\phi = 0.75$ and 0.70 produced solid granules upon high-stress mixing, which remained solid but grew in size after subsequent low-stress mixing. Samples with $\phi = 0.65$ and 0.60 also produced solid granules after high-stress mixing, but these merged to give flowing suspensions after low-stress mixing. Agitating the flowing suspensions under high shear regenerated the solid granules. Samples with $\phi = 0.55$ and 0.50 always produced flowing suspensions.

We next characterised the rheology of the flowing suspensions produced in the range $0.3 \lesssim \phi \lesssim 0.65$.  These samples were prepared from granules made in the high shear mixer at $\phi = 0.65$, which were then vortex mixed to a flowing suspension and diluted to the required volume fraction. These suspensions shear thicken, \autoref{fig:rheo} (inset): the viscosity increases from a lower to a higher Newtonian plateau as stress increases, with the onset stress for thickening, $\sigma^{\star} \approx \SI{1}{\pascal}$, being approximately independent of $\phi$. The viscosity at a fixed stress, $\eta(\sigma;\phi)$, diverges at some $\phi_{\textrm{J}}(\sigma)$. \autoref{fig:rheo} shows the two limiting plots, $\eta_{\textrm{L}}(\phi)$ for the low-stress ($\sigma \ll \sigma^{\star}$) Newtonian plateau and $\eta_{\textrm{H}}(\phi)$ for the high-stress ($\sigma \gg \sigma^{\star}$) Newtonian plateau \footnote{A fully-developed high-$\sigma$ plateau is typically curtailed by sample fracture, evidenced by $\textrm{d}\eta/\textrm{d}\sigma < 0$; we estimate the high-$\sigma$ plateau using the highest viscosity reached.}.
For all $\sigma$, we find that $\eta(\sigma;\phi)$ follows
\begin{equation}
    \eta_\textrm{r}(\sigma;\phi) = \frac{\eta(\sigma;\phi)}{\eta_{\textrm{s}} }= \left[ 1 - \frac{\phi}{\phi_\textrm{J}(\sigma)}\right]^{-\lambda} \label{eq:KD},
\end{equation}
with $\lambda = 1.74 \pm 0.02$. This exercise yields the jamming point as a function of stress, $\phi_{\textrm{J}}(\sigma)$ (Fig.~S3 \cite{SI}). Fitting the limiting branches $\eta_{\textrm{L}}(\phi)$ and $\eta_{\textrm{H}}(\phi)$
returns $\phi_{\textrm{J}}(\sigma \ll \sigma^{\star}) = 0.662$ and $\phi_{\textrm{J}}(\sigma \gg \sigma^{\star}) = 0.568$ respectively.

The rheology shown in \autoref{fig:rheo} is qualitatively identical to the shear thickening behavior of nearly-monodisperse HS \cite{Guy2015,royer2016rheological}, which is caused by the stress-driven transition from lubricated to frictional interparticle contacts \cite{Seto2013,Mari2014,wyart2014,Guy2015,Lin2015,comtet2017pairwise}. At $\sigma \ll \sigma^{\star}$, all contacts are lubricated, and $\eta(\phi)$ diverges at random close packing, $\phi_{\textrm{rcp}}$. On the other hand, when $\sigma \gg \sigma^{\star}$, all contacts are frictional, and $\eta(\phi)$ diverges at a jamming point $\phi_{\textrm{m}} < \phi_{\textrm{rcp}}$. In monodisperse HS, $\phi_{\textrm{rcp}} = 0.64$ and $\phi_{\textrm{m}} = 0.55$ in the limit of large interparticle friction coefficient \cite{Mari2014}. In our polydisperse system, $\phi_{\textrm{rcp}} = 0.662$ and $\phi_{\textrm{m}} = 0.568$.

These two volume fractions demarcate three regimes in \autoref{fig:rheo}. At $\phi<\phi_\textrm{m}$ (red) the system flows at all applied stresses, transitioning from a low- to a high-viscosity state as $\sigma$ increases beyond $\sigma^{\star}$. At $\phi_\textrm{m} \leq \phi  < \phi_\textrm{rcp}$ (yellow) the system flows at $\sigma < \sigma^{\star}$, but jams into a solid-like state \cite{Peters2016,Cates2014} at higher stresses. Finally, at $\phi >\phi_\textrm{rcp}$ (green), there is inadequate liquid to disperse all the particles, and there is no meaningful `suspension rheology'. Significantly, the three types of behavior shown in \autoref{fig:granulation} occur in precisely the three concentration regimes in \autoref{fig:rheo},  pointing to a connection between rheology and liquid incorporation via the physics of jamming \cite{Cates2005,Cates2014}.

To elucidate this connection, we turn to the measured jamming point as a function of applied stress, $\phi_{\textrm{J}}(\sigma)$, which can be fitted well by a stretched exponential \cite{SI}. The inverse function $\sigma_{\textrm{J}}(\phi)$, \autoref{fig:jamming}, gives the `jamming phase boundary' separating flowing and jammed state points.  Red and green regions flow and jam respectively at all stresses, whereas the yellow region will flow at low stresses and form granules at high stress, with the transition between the two states beginning at $\sigma \sim \sigma^{\star}$.

In order to situate our observations, \autoref{fig:granulation}, on \autoref{fig:jamming}, we need to estimate typical stresses encountered in our protocol. The main mechanical action in our high-shear mixer occurs in the $h \approx \SI{1}{\milli\meter}$ gap between the tip of the blade  (speed $v \approx \SI{3}{\meter\per\second}$) and the bowl, at a shear rate of $\sim v/h \sim \SI{3000}{\per\second}$. Multiplying by the appropriate viscosity turns this into a stress. Below $\phi_{\textrm{m}}$, we use the high-stress plateau value; above $\phi_{\textrm{m}}$ we use the high-stress plateau value of the most concentrated sample below $\phi_{\textrm{m}}$ as a lower bound. In our vortex mixer, the centrifugal stress acting on a particle is $\approx \SI{0.2}{\pascal} \lesssim \sigma^{\star}$ \cite{SI}. These crude estimates locate the observations shown in \autoref{fig:granulation} on the jamming phase diagram to within order of magnitude on the stress axis, \autoref{fig:jamming}. It is clear that the jamming phase boundary demarcates granules ($\blacktriangle$) and flowing suspensions ({\color{red} $\blacktriangle$}) in our experiments. Granulation is indeed directly related to jamming.

Cates and others have adumbrated a physical mechanism for this connection \cite{Cates2005,Cates2014}. A suspension droplet at $\phi$ subjected to stress $\sigma > \sigma_{\textrm{J}}(\phi)$ will be in a jammed, solid state. Trying to shear a jammed state at constant $\dot\gamma$ generates very large stresses, which we assume exceed $\sigma_{\textrm{cap}} \sim \Sigma/a$, a capillary stress scale (where $\Sigma$ is the interfacial tension and $a=d/2$ is the particle radius). Particles will therefore protrude from the droplet-air interface. If $\sigma_{\textrm{cap}} > \sigma_{\textrm{J}}(\phi)$, then upon removal of the applied stress, the negatively curved interfaces between protruding particles continue to exert a high enough stress on the droplet to trap it in a jammed, solid state -- it is a granule, which is matt because of protruding particles. Note, in passing, that there has long been experimental evidence for a critical $\Sigma/a$ ratio for granulation \cite{Capes1965}.

Such granules are `fragile'~\cite{Cates1998}: each offers solid-like resistance only to the particular configuration of stresses that jammed it in the first place. Even very low stresses in other directions will unjam the state \cite{Lin2016}. Thus, a granule with $\phi < \phi_{\textrm{rcp}}$ subjected to random stresses below $\sigma_{\textrm{J}}(\phi)$ will unjam and remain fluid, because the external stresses are now insufficient to re-jam it. A collection of such droplets will merge into a bulk, flowing suspension.

To confront observations, note that for us, $\Sigma \lesssim \SI{65}{\milli\newton\per\meter}$ \cite{takamura2012physical} and $a \approx \SI{3}{\micro\meter}$, so that $\sigma_{\textrm{cap}} \sim \Sigma/a \sim \SI{e4}{\pascal}$. Indeed, in all granulated cases, $\sigma_{\textrm{cap}} > \sigma_{\textrm{J}}(\phi)$, \autoref{fig:jamming}, providing direct confirmation of the link between granulation and shear-induced jamming \cite{Cates2005,Cates2014}.


As in a conventional phase diagram, a `lever rule' applies to \autoref{fig:jamming}, in both cases reflecting mass conservation. Consider a mixture with average composition $\phi = \phi_0$ (point A) consisting initially of dry powder at $\phi = 1$ (point C) and liquid sprayed into the mixer at $\phi = 0$. As mixing progresses, particles are incorporated into droplets (all $\gg d$) until their composition reach point B, which for $\sigma \gg \sigma^{\star}$ will be at $\approx \phi_{\textrm{m}}$. Here the droplets jam solid and will not incorporate any more particles, so that some unincorporated dry powder remains. Mass conservation then predicts that the proportion of granules to dry powder is given by the lever-rule construction, viz., AC/AB. Similarly, granulating a mixture starting at state point D gives granules at $\phi_{\textrm{rcp}}$ (point E) coexisting with powder (point F) in proportion DF/DE.

We observed only matt granules but no loose, dry powder up to $\phi \lesssim 0.85$. The most parsimonious assumption is to postulate granules in which the outmost layer of particles protrude, which can be treated as a layer of dry powder. Based on the lever rule (= mass conservation), we can predict the radius of such granules. Working in more general terms for later use, the volume of solid and liquid in a granule of radius $R$ with a dry powder shell of thickness $t_{\textrm{s}}$, which in general may be dependent on stress, and $\phi = 1$ (dry = no liquid) and a jammed core ($\phi = \phi_{\textrm{J}}$) are
\begin{eqnarray}
  V_\textrm{solid} &=& \phi_\textrm{J}(\sigma) \left( \frac{4}{3}\pi R^3\right),\\
  V_\textrm{liquid} &=& [1-\phi_\textrm{J}(\sigma)]\left( \frac{4}{3}\pi [R-t_\textrm{s}(\sigma)]^3\right)
\end{eqnarray}
respectively. Since $\phi = {V_\textrm{solid}}/\left( {V_\textrm{solid} + V_\textrm{liquid}} \right)$, we find
\begin{equation}
R = \frac{t_{\textrm{s}}(\sigma)}{ 1 - \left[  \left(   \frac{\phi_{\textrm{J}}(\sigma)}{1 - \phi_{\textrm{J}}(\sigma)} \frac{1 - \phi}{\phi}         \right)^{1/3}    \right]} . \label{eq:size}
\end{equation}
The predictions for $t_{\textrm{s}} = d$ and $\sigma_{\textrm{J}} = \phi_{\textrm{rcp}}$ and $\phi_{\textrm{m}}$ are shown in \autoref{fig:size} (dot-dash). In both cases, $R$ decreases with $\phi$, starting from a divergence at $\phi_{\textrm{J}}$. The latter is qualitatively consistent with previous reports of diverging granule size `at 100\% liquid saturation' \cite{kristensen1996,ritala1988influence}.

To test \autoref{eq:size} quantitatively, we measured the volume-weighted mean granule size, $\bar{R}(\phi)$, prepared using a smaller version of the high-shear mixer used for the experiments reported so far (to conserve material) \cite{SI}. Our results for granulating at high and low stresses are plotted in \autoref{fig:size} (points). Note that  experiments at high stress below $\phi_{\textrm{rcp}}$ proved impractical: granules fluidised into a high viscosity suspension in the low-stress regions of our instrument, which required a higher torque to flow than this mixer could generate. Nevertheless, $R$ does indeed increases with decreasing $\phi$, with clear evidence for a divergence for the low-$\sigma$ data at $\approx \phi_{\textrm{rcp}}$. The predictions of \autoref{eq:size} capture the form of the experimental data, but are numerically two orders of magnitude too small.

To investigate the cause of this discrepancy we cut open a granule \cite{SI}, revealing a wet core and an essentially dry shell that was many particles thick \footnote{Homogeneous granules in which all particles are slightly wetted and cohere by capillary bridges could be envisaged, but are not observed under our conditions.}.  Such core-shell granule have been observed before \cite{mort2005scale,hapgood2000nucleation}. Presumably there is a very small amount of liquid in this dry shell forming capillary bridges to bind the particles together \cite{Salman2006,Iveson2001a}; but the liquid content is too small to visualise, justifying the assumption of $\phi = 1$ in the shell we used in arriving at \autoref{eq:size}. The assumption that the jammed core is at $\phi_{\textrm{J}}$ is justified by X-ray tomography \cite{SI}, which detects $\lesssim \SI{2}{\percent}_\textrm{vol.}$ air inside granules.

We now fit \autoref{eq:size} to our data, with the shell thickness $t_{\textrm{s}}(\sigma)$ as a stress-dependent parameter, \autoref{fig:size} (dotted). This returns shell thicknesses of $74d$ and $54d$ for high- and low-$\sigma$ respectively, possibly suggesting that high-stress mixing makes available somewhat more liquid for capillary bridges to build a thicker dry shell. Fitting our two data sets using a single shell thickness returns $t_{\textrm{s}} = 60d$, while measurement of a single granule \cite{SI} gave $\approx 45d$. (Multiple measurements failed because granules typically disintegrated during sectioning.) Performing a similar analysis using a polydisperse granule distribution \cite{SI} does not materially change these conclusions.


To summarise, we have shown that the phenomenology of liquid incorporation into powders with repulsive interactions is dominated by shear thickening, which, {\it in extremis}, leads to jamming. Thus, the main control parameter is the stress to which the system is subjected during mixing, because there is a volume-fraction-dependent stress, $\sigma_\textrm{J}(\phi)$, above which frictional contacts result in jamming. Mixing below $\sigma_\textrm{J}(\phi)$ yields flowing suspensions. Mixing above $\sigma_\textrm{J}(\phi)$ yields wet, jammed homogeneous cores coexisting with dry powder. We observed these coexisting states as core and shell respectively in heterogeneous granules. Conservation of mass leads to the prediction of divergent granule sizes as $\phi \to \phi_{\textrm{J}}$ from above, as observed here and in previous work. Quantitatively, a stress-dependent shell size gives good fits to the observed granule size as a function of $\phi$.

Mixing stress has not been identified as a key variable before in granulation: a well-known `granule regime map' \cite{regime} uses the `maximum pore saturation' and a `deformation number' as control parameters. The former roughly plays the role of $\phi^{-1}$, while the latter measures kinetic energy density inside the mixer. These variables control kinetic processes, and affect, e.g., the thickness of the dry `shell' on core-shell granules.

Our incorporation phase diagram, \autoref{fig:jamming}, suggests novel design principles. For example, the formulation space in which granulation can occur could be tuned by varying $\phi_{\textrm{rcp}}$ and $\phi_{\textrm{m}}$ through modification of e.g., the particle size or shape distributions. Separately,  while the outside of a core-shell granule will always be at $\phi = 1$, the density of the core can be `tuned' by moving along $\sigma_\textrm{J}(\phi)$ by using different mixing stresses.
Furthermore, advances in shear-thickening rheology tell us how to modify the $\sigma_\textrm{J}(\phi)$ curve itself: e.g., $\phi_\textrm{m}$ can be tuned by changing the interparticle friction coefficient \cite{Seto2013,Mari2014}, e.g., by using a variety of surface additives \cite{Isa2015}. A full understanding, of course, has to await future work that brings together our incorporation phase diagram and the kinetic regime map \cite{regime}. Note that the kinetic energy density axis in the latter has the dimensions of stress, which is an axis in the former. A deeper connection seems probable.

Finally, many systems exhibit more complex rheology such as shear thinning when additional constraints, e.g., adhesion, are considered \cite{Guy2018constraint}.  The impact of such complexity on the incorporation of liquid into powders is not immediately clear and should be probed in future work.

\begin{methods}
We used soda-lime Potters Spheriglass (A-Type 5000, dried at \SI{120}{\degree C} for 3 hours), which  consists mostly of polydisperse hard spheres (HS; mean diameter $d=\SI{7.2}{\micro m}$, polydispersity = 147\%), but with some irregular shards \cite{SI}. The powder was dispersed into a 9:1 by volume glycerol-water mixture (viscosity $\eta_{\textrm{s}} =  \SI{0.336}{\pascal\second}$) using a two-stepped protocol at various solid volume fractions $\phi = V_{\textrm{solid}}/(V_{\textrm{solid}} + V_{\textrm{liquid}})$ (so that in a dry powder $\phi = 1$) calculated from mass fractions \cite{SI}. High stress mixing using a bespoke mixer is followed by lower stress mixing using an Ika Vortex Genius 3 mixer.

Our bespoke device consisted of a high-torque overhead mixer (Ika Eurostar Power Control-Visc) driving an aluminium impeller with three equi-spaced blades with a \SI{45}{\degree} rake at \SI{500}{\rpm}. Liquid drops (diameter \SI{4.57(7)}{\mm}~$\gg d$) were added at \SI{15}{\ml\per\minute} using a syringe pump (New Era Pump Systems Inc.~NE-1000). Mixing occurred in a glass cylinders with diameter \SI{110}{mm} (\SI{75}{\cm\cubed} batches) or \SI{75}{mm} (\SI{45}{\cm\cubed} batches)\cite{SI}. Rheology measurements were made using an Anton-Paar MCR-301 stress-controlled rheometer with \SI{50}{\milli\meter} roughened parallel plates (surface roughness $\approx \SI{50}{\um}$) separated by \SI{1}{\milli\meter}.  Experiments were performed at \SI{19}{\celsius}.

\end{methods}


\begin{thebibliography}{10}
\expandafter\ifx\csname url\endcsname\relax
  \def\url#1{\texttt{#1}}\fi
\expandafter\ifx\csname urlprefix\endcsname\relax\def\urlprefix{URL }\fi
\providecommand{\bibinfo}[2]{#2}
\providecommand{\eprint}[2][]{\url{#2}}

\bibitem{Iveson2001}
\bibinfo{author}{Iveson, S.~M.}, \bibinfo{author}{Litster, J.~D.},
  \bibinfo{author}{Hapgood, K.~P.} \& \bibinfo{author}{Ennis, B.}
\newblock \bibinfo{title}{{Nucleation, growth and breakage phenomena in
  agitated wet granulation processes: a review}}.
\newblock \emph{\bibinfo{journal}{Powder Tech.}}
  \textbf{\bibinfo{volume}{117}}, \bibinfo{pages}{3--39}
  (\bibinfo{year}{2001}).

\bibitem{Salman2006}
\bibinfo{editor}{Salman, A.~D.}, \bibinfo{editor}{Hounslow, M.} \&
  \bibinfo{editor}{Seville, J.~P.} (eds.) \emph{\bibinfo{title}{Granulation}},
  vol.~\bibinfo{volume}{11} of \emph{\bibinfo{series}{Handbook of Powder
  Technology}} (\bibinfo{publisher}{Elsevier}, \bibinfo{year}{2006}).

\bibitem{marvin1875}
\bibinfo{author}{Marvin, J.}
\newblock \emph{\bibinfo{title}{The Theory and Practice of Granulating
  Gunpowder}} (\bibinfo{publisher}{U.S. Naval Experimental Battery},
  \bibinfo{address}{Annapolis, MD.}, \bibinfo{year}{1875}).

\bibitem{regime}
\bibinfo{author}{Iveson, S.~M.} \& \bibinfo{author}{Lister, J.~D.}
\newblock \bibinfo{title}{Growth regime map for liquid-bound granules}.
\newblock \emph{\bibinfo{journal}{AIChEJ}} \textbf{\bibinfo{volume}{44}},
  \bibinfo{pages}{1510--1518} (\bibinfo{year}{1998}).

\bibitem{blanco2019conching}
\bibinfo{author}{Blanco, E.} \emph{et~al.}
\newblock \bibinfo{title}{Conching chocolate is a prototypical transition from
  frictionally jammed solid to flowable suspension with maximal solid content}.
\newblock \emph{\bibinfo{journal}{Proceedings of the National Academy of
  Sciences}} \textbf{\bibinfo{volume}{116}}, \bibinfo{pages}{10303--10308}
  (\bibinfo{year}{2019}).

\bibitem{SI}
\bibinfo{author}{Hodgson, D. J.~M.}, \bibinfo{author}{Hermes, M.},
  \bibinfo{author}{Blanco, E.} \& \bibinfo{author}{Poon, W. C.~K.}
\newblock \bibinfo{note}{Supplementary Information}.

\bibitem{cazacliu2009concrete}
\bibinfo{author}{Cazacliu, B.} \& \bibinfo{author}{Roquet, N.}
\newblock \bibinfo{title}{Concrete mixing kinetics by means of power
  measurement}.
\newblock \emph{\bibinfo{journal}{Cement and Concrete Research}}
  \textbf{\bibinfo{volume}{39}}, \bibinfo{pages}{182--194}
  (\bibinfo{year}{2009}).

\bibitem{Cates2005}
\bibinfo{author}{Cates, M.~E.}, \bibinfo{author}{Haw, M.~D.} \&
  \bibinfo{author}{Holmes, C.~B.}
\newblock \bibinfo{title}{{Dilatancy, jamming, and the physics of
  granulation}}.
\newblock \emph{\bibinfo{journal}{J. Phys.: Condens. Matter}}
  \textbf{\bibinfo{volume}{17}}, \bibinfo{pages}{S2517--S2531}
  (\bibinfo{year}{2005}).

\bibitem{Cates2014}
\bibinfo{author}{Cates, M.~E.} \& \bibinfo{author}{Wyart, M.}
\newblock \bibinfo{title}{{Granulation and bistability in non-Brownian
  suspensions}}.
\newblock \emph{\bibinfo{journal}{Rheol. Acta}} \textbf{\bibinfo{volume}{53}},
  \bibinfo{pages}{755--764} (\bibinfo{year}{2014}).

\bibitem{Seto2013}
\bibinfo{author}{Seto, R.}, \bibinfo{author}{Mari, R.},
  \bibinfo{author}{Morris, J.~F.} \& \bibinfo{author}{Denn, M.~M.}
\newblock \bibinfo{title}{{Discontinuous Shear Thickening of Frictional
  Hard-Sphere Suspensions}}.
\newblock \emph{\bibinfo{journal}{Phys. Rev. Lett.}}
  \textbf{\bibinfo{volume}{111}}, \bibinfo{pages}{218301}
  (\bibinfo{year}{2013}).

\bibitem{Mari2014}
\bibinfo{author}{Mari, R.}, \bibinfo{author}{Seto, R.},
  \bibinfo{author}{Morris, J.~F.} \& \bibinfo{author}{Denn, M.~M.}
\newblock \bibinfo{title}{{Shear thickening, frictionless and frictional
  rheologies in non-Brownian suspensions}}.
\newblock \emph{\bibinfo{journal}{J. Rheol.}} \textbf{\bibinfo{volume}{58}},
  \bibinfo{pages}{1693--1724} (\bibinfo{year}{2014}).

\bibitem{wyart2014}
\bibinfo{author}{Wyart, M.} \& \bibinfo{author}{Cates, M.~E.}
\newblock \bibinfo{title}{Discontinuous shear thickening without inertia in
  dense non-brownian suspensions}.
\newblock \emph{\bibinfo{journal}{Phys. Rev. Lett.}}
  \textbf{\bibinfo{volume}{112}}, \bibinfo{pages}{098302}
  (\bibinfo{year}{2014}).

\bibitem{Guy2015}
\bibinfo{author}{Guy, B.~M.}, \bibinfo{author}{Hermes, M.} \&
  \bibinfo{author}{Poon, W. C.~K.}
\newblock \bibinfo{title}{{Towards a Unified Description of the Rheology of
  Hard-Particle Suspensions}}.
\newblock \emph{\bibinfo{journal}{Phys. Rev. Lett.}}
  \textbf{\bibinfo{volume}{115}}, \bibinfo{pages}{088304}
  (\bibinfo{year}{2015}).

\bibitem{Lin2015}
\bibinfo{author}{Lin, N. Y.~C.} \emph{et~al.}
\newblock \bibinfo{title}{Hydrodynamic and contact contributions to continuous
  shear thickening in colloidal suspensions}.
\newblock \emph{\bibinfo{journal}{Phys. Rev. Lett.}}
  \textbf{\bibinfo{volume}{115}}, \bibinfo{pages}{228304}
  (\bibinfo{year}{2015}).

\bibitem{comtet2017pairwise}
\bibinfo{author}{Comtet, J.} \emph{et~al.}
\newblock \bibinfo{title}{Pairwise frictional profile between particles
  determines discontinuous shear thickening transition in non-colloidal
  suspensions}.
\newblock \emph{\bibinfo{journal}{Nat. Commun.}} \textbf{\bibinfo{volume}{8}}
  (\bibinfo{year}{2017}).

\bibitem{royer2016rheological}
\bibinfo{author}{Royer, J.~R.}, \bibinfo{author}{Blair, D.~L.} \&
  \bibinfo{author}{Hudson, S.~D.}
\newblock \bibinfo{title}{Rheological signature of frictional interactions in
  shear thickening suspensions}.
\newblock \emph{\bibinfo{journal}{Phys. Rev. Lett.}}
  \textbf{\bibinfo{volume}{116}}, \bibinfo{pages}{188301}
  (\bibinfo{year}{2016}).

\bibitem{Peters2016}
\bibinfo{author}{Peters, F.} \emph{et~al.}
\newblock \bibinfo{title}{Rheology of non-brownian suspensions of rough
  frictional particles under shear reversal: A numerical study}.
\newblock \emph{\bibinfo{journal}{J. Rheol.}} \textbf{\bibinfo{volume}{60}},
  \bibinfo{pages}{715--732} (\bibinfo{year}{2016}).

\bibitem{Capes1965}
\bibinfo{author}{Capes, C.~E.} \& \bibinfo{author}{Danckwerts, P.~V.}
\newblock \bibinfo{title}{Granule formation by the agglomeration of damp
  powders: 1. the mechanism of granule growth}.
\newblock \emph{\bibinfo{journal}{Trans. Inst. Chem. Eng.}}
  \textbf{\bibinfo{volume}{43}}, \bibinfo{pages}{116--124}
  (\bibinfo{year}{1965}).

\bibitem{Cates1998}
\bibinfo{author}{Cates, M.~E.}, \bibinfo{author}{Wittmer, J.~P.},
  \bibinfo{author}{Bouchaud, J.-P.} \& \bibinfo{author}{Claudin, P.}
\newblock \bibinfo{title}{Jamming, force chains, and fragile matter}.
\newblock \emph{\bibinfo{journal}{Phys. Rev. Lett.}}
  \textbf{\bibinfo{volume}{81}}, \bibinfo{pages}{1841--1844}
  (\bibinfo{year}{1998}).

\bibitem{Lin2016}
\bibinfo{author}{Lin, N.~Y.}, \bibinfo{author}{Ness, C.},
  \bibinfo{author}{Cates, M.~E.}, \bibinfo{author}{Sun, J.} \&
  \bibinfo{author}{Cohen, I.}
\newblock \bibinfo{title}{Tunable shear thickening in suspensions}.
\newblock \emph{\bibinfo{journal}{Proc. Natl. Acad. Sci. (USA)}}
  \textbf{\bibinfo{volume}{113}}, \bibinfo{pages}{10774--10778}
  (\bibinfo{year}{2016}).

\bibitem{takamura2012physical}
\bibinfo{author}{Takamura, K.}, \bibinfo{author}{Fischer, H.} \&
  \bibinfo{author}{Morrow, N.~R.}
\newblock \bibinfo{title}{Physical properties of aqueous glycerol solutions}.
\newblock \emph{\bibinfo{journal}{Journal of Petroleum Science and
  Engineering}} \textbf{\bibinfo{volume}{98}}, \bibinfo{pages}{50--60}
  (\bibinfo{year}{2012}).

\bibitem{kristensen1996}
\bibinfo{author}{Kristensen, H.~G.}
\newblock \bibinfo{title}{Particle agglomeration in high shear mixers}.
\newblock \emph{\bibinfo{journal}{Powder Tech.}} \textbf{\bibinfo{volume}{88}},
  \bibinfo{pages}{197--202} (\bibinfo{year}{1996}).

\bibitem{ritala1988influence}
\bibinfo{author}{Ritala, M.}, \bibinfo{author}{Holm, P.},
  \bibinfo{author}{Schaefer, T.} \& \bibinfo{author}{Kristensen, H.}
\newblock \bibinfo{title}{Influence of liquid bonding strength on power
  consumption during granulation in a high shear mixer}.
\newblock \emph{\bibinfo{journal}{Drug Dev. Ind. Pharm.}}
  \textbf{\bibinfo{volume}{14}}, \bibinfo{pages}{1041--1060}
  (\bibinfo{year}{1988}).

\bibitem{mort2005scale}
\bibinfo{author}{Mort, P.~R.}
\newblock \bibinfo{title}{Scale-up of binder agglomeration processes}.
\newblock \emph{\bibinfo{journal}{Powder Technology}}
  \textbf{\bibinfo{volume}{150}}, \bibinfo{pages}{86--103}
  (\bibinfo{year}{2005}).

\bibitem{hapgood2000nucleation}
\bibinfo{author}{Hapgood, K.~P.}
\newblock \emph{\bibinfo{title}{Nucleation and binder dispersion in wet
  granulation}}.
\newblock Ph.D. thesis, \bibinfo{school}{Department of Chemical Engineering,
  University of Queensland} (\bibinfo{year}{2000}).

\bibitem{Iveson2001a}
\bibinfo{author}{Iveson, S.} \emph{et~al.}
\newblock \bibinfo{title}{{Growth regime map for liquid-bound granules: further
  development and experimental validation}}.
\newblock \emph{\bibinfo{journal}{Powder Tech.}}
  \textbf{\bibinfo{volume}{117}}, \bibinfo{pages}{83--97}
  (\bibinfo{year}{2001}).

\bibitem{Isa2015}
\bibinfo{author}{Fernandez, N.}, \bibinfo{author}{Cayer-Barrioz, J.},
  \bibinfo{author}{Isa, L.} \& \bibinfo{author}{Spencer, N.~D.}
\newblock \bibinfo{title}{Direct, robust technique for the measurement of
  friction between microspheres}.
\newblock \emph{\bibinfo{journal}{Langmuir}} \textbf{\bibinfo{volume}{31}},
  \bibinfo{pages}{8809--8817} (\bibinfo{year}{2015}).

\bibitem{Guy2018constraint}
\bibinfo{author}{Guy, B.~M.}, \bibinfo{author}{Richards, J.~A.},
  \bibinfo{author}{Hodgson, D. J.~M.}, \bibinfo{author}{Blanco, E.} \&
  \bibinfo{author}{Poon, W. C.~K.}
\newblock \bibinfo{title}{Constraint-based approach to granular dispersion
  rheology}.
\newblock \emph{\bibinfo{journal}{Phys. Rev. Lett.}}
  \textbf{\bibinfo{volume}{121}}, \bibinfo{pages}{128001}
  (\bibinfo{year}{2018}).

\end{thebibliography}


\begin{addendum}
 \item We thank Fran\c cois Lequeux (ESPCI, Paris) for directing us to reference \cite{regime}. DH held an EPSRC studentship. EB was funded by Mars Chocolates UK Ltd. MH and WP were funded by EPSRC grants EP/J007404/1 and EP/N025318/1.
 \item[Competing Interests] The authors declare that they have no
competing financial interests.
 \item[Correspondence] Correspondence and requests for materials
should be addressed to Daniel J. M. Hodgson~(email: danieljmhodgson@gmail.com).
 \item[Data availability] The data in this work are available for download at \url{https://doi.org/10.7488/ds/2588}.
\end{addendum}

\end{document}